**A tale of two species: How humans shape dog behaviour in urban habitats**

Debottam Bhattacharjee[1], Rohan Sarkar[1], Shubhra Sau[1] and Anindita Bhadra[1*]

**Affiliation**:

[1] Department of Biological Sciences, Indian Institute of Science Education and Research Kolkata, Nadia, West Bengal, India

[*]**Address for Correspondence**

Behaviour and Ecology Lab, Department of Biological Sciences,

Indian Institute of Science Education and Research Kolkata

Mohanpur Campus, Mohanpur, Nadia

PIN 741246, West Bengal, INDIA

[*]Corresponding author

E-mail: abhadra@iiserkol.ac.in (AB)

**Abstract**

Species inhabiting urban environments experience enormous anthropogenic stress. Behavioural plasticity and flexibility of temperament are crucial to successful urban-adaptation. Urban free-ranging dogs experience variable human impact, from positive to negative and represent an ideal system to evaluate the effects of human-induced stress on behavioural plasticity. We tested 600 adult dogs from 60 sites across India, categorised as high (HF), low (LF), and intermediate (IF) human flux zones, to understand their sociability towards an unfamiliar human. Dogs in the HF and IF zones were more 'bold' and as compared to their 'shy' counterparts in LF zones. The IF zone dogs were the most sociable. This is the first-ever study aimed to understand how an animal's experiences of interactions with humans in its immediate (natural) environment might shape it's response to humans. This is very relevant in the context of human-animal conflict induced by rapid urbanization and habitat loss across the world.

Rapid urbanization across the world has led to emergence of a wide range of urban-adapted species that have the ability to survive in human-dominated urban ecosystems. Urban-adaptation entails modifications in the behavioural repertoire of species. Behavioural plasticity and/or change in temperament in response to urbanization can improve an organism's chances of survival and reproduction in an altered, often hostile, environment[1]. Altered foraging behaviour of hedgehogs to avoid crowded areas in daylight[2], the higher pitch in the calls of great tits in a noisy environment[3,4], bolder and more exploratory behaviour of urban coyotes[5] etc. are some examples of behavioural alterations of urban-adapted species.

Urbanised populations display 'bold' behavioural responses as compared to their 'shy' rural counterparts[6,7]. Some recent studies exploring the genetic basis for behavioural flexibility in urbanised species have suggested micro-evolutionary mechanisms to be responsible for such adaptation[8–10]. For some species, an additional dimension of complexity prevails in their behavioural traits, enabling them to adapt to and interact with humans directly or indirectly[11]. Urban free-ranging dogs (*Canis lupus familiaris*) are an example of urban-adapted species that have co-evolved with humans. These dogs not only depend on humans for food[12], and shelter[13], but receive a range of negative interactions including beating, threatening, harassment, and poisoning as well[14]. Surprisingly, human induced stress or impact on the eco-ethology of dogs in the urban environment has never been assessed adequately except for a handful of studies that explored the attributes of dog-human relationship on the streets[14].

Sociability is one of many personality traits in animals that is often considered within the continuum of a shyness – boldness axis, and is also observed from the perspective of social fearfulness and aggression[15–19]. Studies have explored the change in temperament and behavioural differences induced by stress in pet and shelter dogs[20–22]. However, there is a lacuna in our understanding of the negative effects of human interactions on the sociability of dogs in a natural habitat, devoid of caring humans. In this study, we aimed to investigate the sociability of Indian free-ranging dogs to identify potential drivers for changes in temperament or behavioural flexibility.

We used human flux as a surrogate of human-induced stress on the urban free-ranging dogs. We categorized all sampled areas into three human flux zones (high, intermediate, and low) and tested dogs for sociability. We hypothesized that sociability would be the least in high human flux zones and the highest in low flux zones. Additionally, we expected that resource abundance of an area would predict the density of dogs, but not the sociability trait. In a broader sense, we assume that any change in dogs' temperament and subsequent behavioural plasticity would be a translation of human-induced stress and a crucial requirement for them to survive in a human-dominated environment.

**Results**

*Characterization of zones:* Human flux was found to vary between the three zones (Kruskal-Wallis test: $\chi^2$ = 52.510, df = 2, p < 0.0001). The HF zones (flux: 81.35 ±11.98) showed the highest flux compared to both IF (flux: 29.1 ± 4.31, Mann-Whitney U test: U = 400.000, df1 = 20, df2 = 20, p < 0.0001) and LF (flux: 5.9 ± 2.7, Mann-Whitney U test: U = 400.000, df1 = 20, df2 = 20, p < 0.0001) zones. Additionally, IF zones had a higher flux compared to LF zones (Mann-Whitney U test: U = 400.000, df1 = 20, df2 = 20, p < 0.0001).

*Approach rate:* 75.5%, 78.5% and 34.5% of the dogs approached the human experimenter in the HF, IF, and LF zones respectively, irrespective of the PV and S phases. There was significant variation in the overall response rates, considering all the three zones (Goodness of fit: $\chi^2$ = 38.472, df = 2, p < 0.0001). Dogs in both the HF and IF zones showed similar positive response rates (Goodness of fit: $\chi^2$ = 0.117, df = 1, p = 0.732), which were higher than in the LF zones (HF - LF; Goodness of fit: $\chi^2 = 30.564$, df = 1, p < 0.0001; IF – LF; Goodness of fit: $\chi^2 = 34.265$, df = 1, p < 0.0001).

Phase-specific comparisons (PV vs S) provided interesting insights into the tendency of dogs to approach the experimenter. The level of positive response varied significantly between the PV and S phases of the three human flux zones (Contingency $\chi^2$: $\chi^2 = 103.917$, df = 2, p < 0.0001, Fig 1). Significantly higher number of approaches were noticed in the S phase compared to the PV phase in HF and LF zones (HF - Goodness of fit: $\chi^2 = 87.583$, df = 1, p < 0.0001; LF - Goodness of fit: $\chi^2 = 17.892$, df = 1, p < 0.001). However, we found significantly higher number of approaches in the PV phase compared to the S phase in the IF zones (Goodness of fit: $\chi^2 = 15.783$, df = 1, p < 0.0001). Considering cumulative responses, the tendency of dogs to approach was more in the S phase, irrespective of the human flux zones (Goodness of fit: $\chi^2 = 23.939$, df = 1, p < 0.0001).

*Latency:* The latency to approach the experimenter varied in the three human flux zones (Kruskal-Wallis test: $\chi^2 = 40.596$, df = 2, p < 0.0001). Dogs in the IF zones approached faster in comparison to HF (Mann-Whitney U test: U = 16673.000, df1 = 151, df2 = 157, p < 0.0001) and LF (Mann-Whitney U test: U = 7043.500, df1 = 157, df2 = 69, p < 0.0001) zones. We did not find any difference between the latencies of HF and LF zone dogs (Mann-Whitney U test: U = 5802.500, df1 = 151, df2 = 69, p = 0.176).

*Demeanour:* The dogs in the three human flux zones showed significant variation in their demeanour towards the experimenter (Contingency $\chi^2$: $\chi^2 = 64.310$, df = 6, p < 0.0001, Fig 2). Demeanours were more or less similar in the HF and IF zones where majority of the dogs showed affiliative (HF: 37.5% and IF: 41%) and anxious (HF: 41.5% and IF: 35.5%) behavioural states. Only 20% and 23% of the dogs displayed a neutral attitude towards the human experimenter in the HF and IF zones respectively. However, a significantly higher percentage of dogs (71%) showed an anxious demeanour, as compared to other behavioural states in the LF zones (Goodness of fit: $\chi^2 = 35.280$, df = 1, p < 0.0001). Aggression was absent or very limited in all the zones (total of 3 dogs displayed aggression in HF and IF zones, none of the dogs in LF zones were aggressive).

*Effects of demeanour and zones on approach:* We investigated the effects of demeanour and zones on the overall approach (irrespective of phases) response using a GLM analysis with Binomial distribution (Supplementary Table 1). LF zones predicted significantly higher 'no approach' but we did not see any effect of demeanour. Further analysis considering responses only in the PV phase revealed a significant effect of demeanour. We found that 71% of the dogs displayed affiliative responses, which was higher than the other demeanours clubbed together (Goodness of fit: $\chi^2 = 24.688$, df = 1, p < 0.0001).

*Affiliation Index (AI):* The AI values of dogs showed significant variation between the three human flux zones (Kruskal-Wallis test: $\chi^2 = 109.554$, df = 2, p < 0.0001, Fig 3). Dogs in the IF zones showed higher AI values compared to both the HF (Mann-Whitney U test: U = 25086.000, df1 = 200, df2 = 200, p < 0.0001) and LF (Mann-Whitney U test: U = 31042.500, df1 = 200, df2 = 200, p < 0.0001) zones, suggesting a higher tendency to respond in a friendly manner. We also found a difference between the HF and LF zone dogs (Mann-Whitney U test: U = 28001.500, df1 = 200, df2 = 200, p < 0.0001), with higher AI values in the HF zones dogs.

*Density of dogs:* The density of dogs, as estimated from the censuses, showed significant variation between the different zones (Kruskal-Wallis test: $\chi^2 = 7.261$, df = 2, p = 0.02, Supplementary Figure 1). Dog density was higher in the HF zones compared to the LF zones (Mann-Whitney U test: U = 298.000, df1 = 20, df2 = 20, p = 0.007). We did not find any variation between HF – IF (Mann-Whitney U test: U = 269.000, df1 = 20, df2 = 20, p = 0.06) and IF – LF (Mann-Whitney U test: U = 220.500, df1 = 20, df2 = 20, p < 0.58) zones.

*Availability of food resources:* RI differed significantly between the three zones (Kruskal-Wallis test: $\chi^2$ = 10.395, df = 2, p = 0.006). Post-hoc pairwise comparisons revealed a difference between HF and LF zones (Mann-Whitney U test: U = 308.500, df1 = 20, df2 = 20, p = 0.003), but no difference between HF – IF (Mann-Whitney U test: U = 283.000, df1 = 20, df2 = 20, p = 0.02) and IF – LF (Mann-Whitney U test: U = 252.000, df1 = 20, df2 = 20, p = 0.16) zones.

The area sampled for each census bout in the three zones was comparable (Kruskal-Wallis test: $\chi^2$ = 1.285, df = 2, p = 0.52, Supplementary Figure 2). We also investigated the correlation between the RI values and corresponding dog abundance in the sampling sites, separately for the three zones (Supplementary Figure 3). We found a moderately strong positive correlation in the HF zones (Spearman Rank Correlation: $r_s$ = 0.47, df = 20, p = 0.03). In the IF and LF zones, we found no correlation between RI values and corresponding dog abundance (IF - Spearman Rank Correlation: $r_s$ = 0.39, df = 20, p = 0.08); LF - Spearman Rank Correlation: $r_s$ = 0.22, df = 20, p = 0.33).

*Human perception:* Comparing the cumulative PI values of the three zones revealed a significant variation (Kruskal-Wallis test: $\chi^2$ = 18.890, df = 2, p < 0.001). Post-hoc pairwise comparisons further disclosed a difference between human perception in the HF and IF zones (Mann-Whitney U test: U = 373.000, df1 = 20, df2 = 20, p < 0.001), with higher PI values in the IF zones. We did not find any difference between HF – LF (Mann-Whitney U test: U = 279.000, df1 = 20, df2 = 20, p = 0.03) and IF – LF (Mann-Whitney U test: U = 252.500, df1 = 20, df2 = 20, p = 0.15) zones.

Apart from calculating the PI, we also documented the reasons cited by people for their perception of free-ranging dogs, which have been provided in Supplementary Table 2.

*Effect of sex on AI:* GLM analysis revealed no effect of sex of the dogs on their corresponding AI values (Supplementary Table 3).

*Effect of zones, RI, cumulative PI on cumulative AI:* GLMM analysis (using Poisson distribution) with human flux zones, RI, cumulative PI as predictors and sampling sites as random factors showed a partial effect on AI (Supplementary Table 4). Only zones were significant predictors and RI and cumulative PI had no effect on AI.

**Discussion**

Our results suggest a strong effect of human flux zones on the sociability trait of urban free-ranging dogs. Dogs in the HF zones were found to be highly opportunistic as they mostly approached the experimenter only when provided with food, while being reluctant to approach as a response to positive vocalization. This clearly advocates the idea that dogs present in busy and crowded urban areas maintain a certain distance from unfamiliar humans, probably to avoid or minimize any unprecedented conflict. A high latency to approach the stranger, further strengthens this conclusion. Since free-ranging dogs often receive negative interactions from humans, such opportunistic behaviour of the dogs in HF zones was not unexpected. However, the largely shy response of the dogs in the LF zones was surprising; since these dogs largely do not encounter unfamiliar humans, we expected them to be less wary of humans than dogs in the HF zones. Interestingly, IF zone dogs were the most bold in their response to the unfamiliar human, approaching him mostly in response to the positive vocalization. These observations lead us to conclude that the dogs in the IF zones mostly have positive experiences with humans, through provisioning of food and shelter and/or petting and are thus less wary of unfamiliar humans. They encounter humans more than the LF but considerably less than HF zones, and mostly encounter familiar humans. This could potentially translate into higher socialization while allowing less window for interspecific conflict with humans at the same time. LF zone dogs, on the other hand, encounter fewer humans on the whole and thereby, have fewer interactions of either kind. Thus, their response to the experimenter is reminiscent of the inherent tendency of any free-ranging animal to avoid humans[2,23,24]. Thus, this study provides strong evidence for the critical role that life experiences play in shaping behaviour and reveals a continuum in the response of dogs to humans in the context of urbanization.

The free-ranging dogs' behavioural plasticity in their sociability level revealed interesting variations along the 'shyness-boldness axis', with the HF and IF zone dogs being 'bolder' compared to their 'shy' counterparts in the LF zones. Varying levels of human socialization and ontogenic experience must have been the key factors of such outcomes. Earlier, shelter dogs were found to display fear-appeasement behaviours compared to pet dogs when confronted with an unfamiliar human, substantiating the role of less socialization and stressful living conditions in shelters[21]. Our study measured similar behavioural differences and change in temperament in dogs using human flux as a proxy for human-induced stress. As hypothesized, we found a moderately strong correlation between available resources and dog abundance in the HF zones, but resource availability could not predict dogs' sociability. Also, we did not obtain a significant prediction from the human perception index. In HF zones, dogs probably undergo enormous pressure of human-induced stress (negative socialization) and higher competition for food and territory, which shape their overall behavioural repertoire and make them well versed in utilizing opportunistic situations, while avoiding direct interactions with humans as far as possible. IF zone dogs most likely receive higher positive human socialization and less competition for food and habitat, as also suggested by the responses in our surveys, eventually translating into affiliative behavioural states. On the other hand, least human socialization in the LF zone dogs might play a bigger role in developing them into 'shy' and hesitant individuals when it comes to facing a novel situation. The fact that these responses were independent of the sex of the dogs suggest a uniform effect of the habitat, emphasizing the roles of socialization and ontogeny. These results imply the sole effect of human flux on the behavioural plasticity in the sociability trait and discard other properties like dog density, available resources and perception of humans.

In conclusion, free-ranging dogs showed varied behavioural traits and personality types at the urban microhabitat levels in response to unknown humans. Versatility of this range in the sociability trait of free-ranging dogs could have been a key factor in their successful colonisation and sustenance in a largely human-dominated habitat and making dogs an 'urban adapted' species. This is also arguably one of the first evidences that animals belonging to the same urban habitat can respond differently, based on different life experiences.

**Methods**

*(i) Study area and categorization of zones -*

In order to characterize the areas with respect to human activity levels, we carried out sampling in some microhabitats like residential areas, busy markets, bus stations, and railway stations in two cities – one small and another very large (Shillong, Meghalaya - 25°57'88"N, 91°89'33"E and Kolkata, West Bengal - 22°57'26"N, 88°36'39"E, India). We randomly chose a spot in the city and stood at the same spot for 1 minute. During this time, we counted the number of people and vehicles that passed the spot (without identifying them). We did such counting multiple times at multiple microhabitats. We gave equal weightage to an individual and a vehicle, as each was counted as a single event of movement. Also, this reduced the chances of over estimation as often it was not possible to correctly estimate the number of people in the vehicle (for example inside a bus). We estimated the human flux in each area as the number of events of movement (pedestrians or vehicles) per minute. We prepared a distribution of the human flux in the different areas and found that it could be divided into three broad zones. Areas like market places, railway stations, bus stations (all non-residential arears) had high human flux (95.22 ± 9.16), which we categorized as high flux zones (HF, Supplementary Figure 4 and 6a). Exclusively residential areas typically had low flux (6.5 ± 1.4) and were categorised as low flux zones (LF, Supplementary Figure 4 and 6b). Intermediate flux zones (IF, Supplementary Figure 4 and 6c) typically had a few shops, eateries, and restaurants within residential areas with an intermediate human flux (20.75 ±4.26). Using this baseline data, we defined the following ranges of the three zones – HF: $\geq$ 60, $60 \leq IF > 10$, $LF \leq 10$.

We carried out the sociability test for free-ranging dogs in nine different cities and their surrounding areas across India (Supplementary Figure 5). The cities were - Kolkata (22°57'26"N, 88°36'39"E), Pune (18°52'04"N, 73°85'67"E), Bengaluru (12°97'16"N, 77°59'46"E), Hyderabad (17°38'50"N, 78°48'67"E), Bhopal (23°25'99"N, 77°41'26"E), Shillong (25°57'88"N, 91°89'33"E), Lucknow (26° 84'67"N, 80°94'62"E), Chandigarh (30°73'33"N, 76°77'94"E), and Jaipur (26°91'24"N, 75°78'73"E). We sampled a cumulative area of 27.33 Sq Km (Google Earth, version 9.2.76.4) in this study, including 60 different and widespread sampling sites, which were further grouped into HF, IF, and LF zones (20 in each category, Supplementary Table 5).

*(ii) Experimental procedure*

*(a) Sociability test* - We quantified the sociability of dogs towards an unfamiliar human experimenter in an obligatory positive vocalisation phase (PV phase), followed by an optional stimulus phase (S phase) in all three zones. S phase was carried out only after an inefficacious PV phase. Language-independent vocal sounds (e.g. "tch tch"). We used a glucose biscuit as a stimulus paired with positive vocalisation in the succeeding S phase, which thus offered a reward to the dog. Protein-rich food such as raw chicken was not used in order to ensure that the effect of the stimulus was not over-amplified.

We tested 200 adult free-ranging dogs from each of the three zones, where the experimenter (E) was constant throughout the experiment and was assisted by a cameraperson. E was male, 165 – 166 cm in height with a slim body structure. The experimental procedure was as follows -

<u>Positive Vocalisation (PV) phase</u> – E approached a random solitary dog in a predefined zone and initially maintained an approximate distance of 1.5 m from the dog. Since the dogs were free-ranging and not on leash, E had to adjust his position with respect to the dog, using eye estimation distance. Immediately after standing at the ideal position, E used positive vocalisations for 5 sec to attract the attention of the dog and waited for a maximum of 25 sec for the dog to respond (Supplementary Movie 1 and 2). E gazed at the individual throughout this phase of the experiment. The whole procedure was recorded by the cameraperson from a minimum distance of 4.5 m, without any intervention in the experimental process. An individual was considered to show approach response when it came within a distance of 0.3 m from E. If the dog did not respond by approaching (> 0.3 m from E), the S phase was carried out.

<u>Stimulus (S) phase</u> – Immediately (2-3 s) after the PV phase, E placed a biscuit (stimulus) on the ground at an approximate distance of 0.1 m from him and roughly 1.4 m away from the dog keeping the total distance consistent with the earlier phase (Supplementary Movie 3). Approach was considered only when an individual obtained the food or inspected it by sniffing or licking or both. This distance was chosen since an individual could simply stand more than 0.3 m away from E and obtain the food if it was placed 0.3 m from E (by stretching his / her neck). Placement of food at a distance of 0.1 m from E compelled individuals to approach E at 0.3 m or less in order to inspect or obtain it, retaining the definition of approach. Positive vocalisations were also used for 5 s while providing the stimulus. Similar to the PV phase, E provided an additional duration of 25 sec to check the response while gazing at the individual for the whole time. The procedure was recorded exactly like earlier. S phase was run for a period of 30 s or ended when an individual inspected or obtained the food, whichever was earlier.

*(b) Quantification of ecological properties of the study area* - We investigated the following ecological properties of the three zones -

<u>Density of dogs</u> – We used the protocol of carrying out dog census standardized by Sen Majumdar et al. 2014[25]. The experimenter randomly walked on the streets of a predefined site to look for free-ranging dogs during 1600 – 1800 hrs. All sightings of dogs, with details about their gender and group size were noted and the locations were recorded using *Google my maps application* in a *One Plus 5T* mobile device. We used Google Earth (version 9.2.76.4) to calculate the area of the sampling site. We calculated the density (individuals / km$^2$) of dogs as using the following formula -

$$\text{Density of dogs} = \frac{\text{Total no. of dogs observed in a sampling site}}{\text{Area of the sampling site}}$$

<u>Availability of food resources</u> – The dog census also included a resource census, in which we identified the potential food sources that dogs could access in all the 60 sampling sites. These included open garbage dumps, eateries, restaurants, meat and fish shops, etc.

Human perception of free-ranging dogs - We constructed a small questionnaire to understand human perception of the free-ranging dogs in the areas where we carried out the census and experiments. We randomly approached 20 people (only adults) from each of the 60 sampling sites and recorded their views while trying not to disclose the purpose of the study to avoid any bias. We asked three generic questions during the survey (Supplementary Table 6).

*(iii) Data analysis and statistics*

We coded the videos and defined the following parameters - approach, no approach, and latency, listed in Supplementary Table 7. Behavioural states or demeanour (holistic) of the individuals were divided into affiliative, neutral, anxious, and aggressive (Please see Bhattacharjee et al. 2018[26] for more details). Shapiro -Wilk tests were carried out to check for normality of data. It was not normally distributed, thus non-parametric tests were used. Generalised linear models (GLM) were performed using "lme4" package of R Studio. Akaike information criterion (AIC) values were compared in order to obtain the best-fitting models and the ones with the lowest values were selected. A second coder, naïve to the purpose of the study, coded 20% of the data to check for reliability. It was perfect for approach (cohen's kappa = 1.00) and almost perfect for latency (cohen's kappa = 0.97). The alpha level was 0.05 throughout the analysis. Post-hoc pairwise comparisons were carried out with Bonferroni correction whenever required and the adjusted p value was considered as 0.016. Besides R Studio, statistical analyses were also performed using StatistiXL (version 1.11.0.0).

We constructed an index, defined as the affiliation index (AI), based on three components – phase-based approach, latency, and demeanour. All the three components were sub-divided into different categories or ranges and scored accordingly (Supplementary Table 8). Scores were assigned to each response on the basis of free-ranging dogs' tendency to socialise with an unfamiliar human. A final cumulative AI value was calculated on a scale of 3 to 12, with the higher values signifying a higher tendency to socialize with the unfamiliar human experimenter.

Apart from the 'Affiliation Index' (AI), we built two additional indices namely 'Resource Index' (RI) and 'Perception Index' (PI). Similar to AI, PI was also built from the perspective of dogs and hence biased towards affiliative / friendly human beings.

Resource Index (RI) – We quantified all the regular and irregular food sources and scored them accordingly keeping in mind the impact each source can have on dogs' nutrition. It is important to note that all the food resources were human generated and we assigned the scores from the perspective of dogs. We have summarised RI in Supplementary Table 9.

RI was constructed in such a way that it could provide the best estimation of potential availability of resources in the 60 sampling sites. It was cumulative in nature i.e. a sampling site would have a cumulative RI value describing its potential resource availability for the free-ranging dogs. For example, a sampling site has only 2 open garbage dumps and 1 meat shop, the RI would be = (2 * 5) + (1 * 8) = 18.

Perception Index (PI) – PI was constructed from the perspective of free-ranging dogs and was a surrogate measure of obtaining human perception in the specific sampling sites. We have summarized the components and assigned scores to construct PI in Supplementary Table 10.

Higher scores determined positive impacts from humans. According to PI, a person who was friendly, affectionate and provisioned food to dogs got a score of 8 and subsequently the lowest score of 3 determined negative influences on dogs.

**Acknowledgements**

This work has been funded by Science and Engineering Research Board (SERB), Department of Science and Technology, Govt. of India (Project No. EMR / 2016 / 000595). R.S and S.S were supported by fellowships from the same project. D.B was supported by a DST-INSPIRE Fellowship. Authors would like to thank Ms. Sudeshna Chakraborty for playing the role of a naïve coder. Authors acknowledge Indian Institute of Science Education and Research Kolkata (IISER-K) for providing the infrastructure.

**Author contributions**

A.B and D.B conceived the study. D.B, R.S, and S.S participated in data acquisition. D.B analysed the data. A.B and D.B interpreted the data and wrote the paper. All authors approved the final manuscript.

**Competing interests**

The authors declare no competing interests.

**Figure and Figure Legends –**

**Fig. 1**

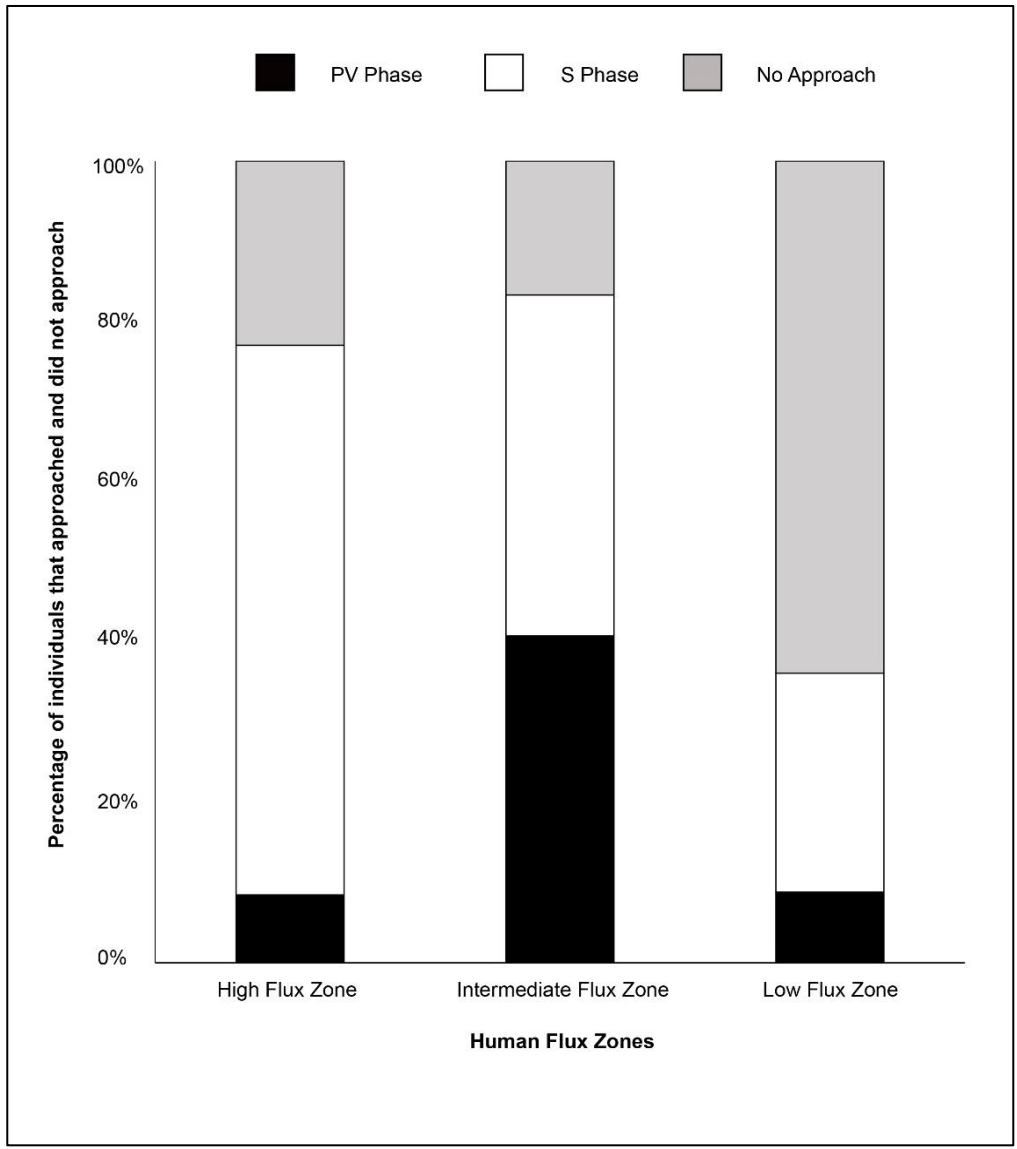

**Fig. 1 Percentage of approach** Stacked bar graph showing the percentages of individuals that approached and did not approach in the PV and S phases of high, intermediate and low human flux zones. The number of individuals that approached in the PV phase were subtracted from the total number of dogs tested for calculating the percentage of approach in the S phase.

**Fig. 2**

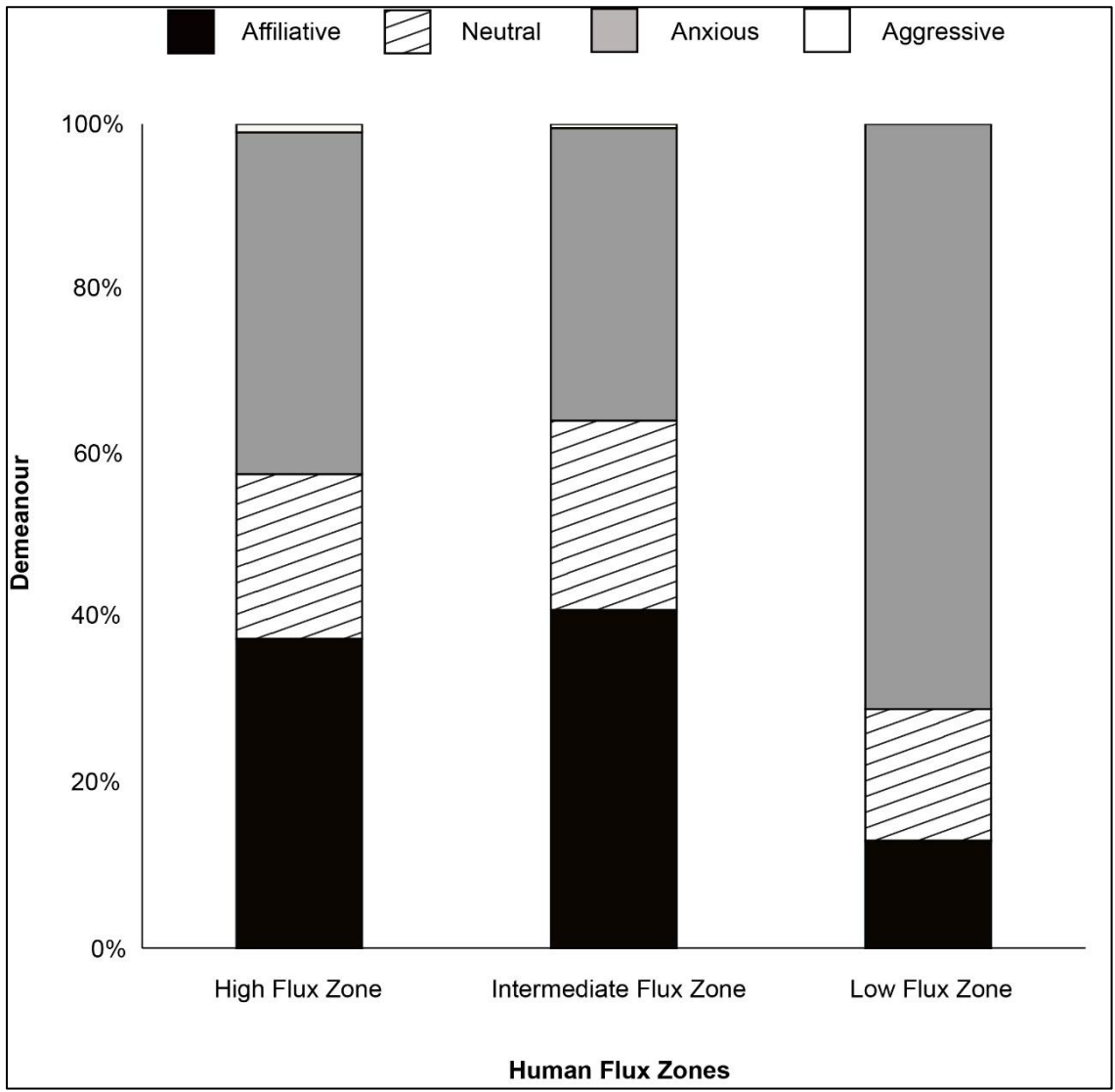

**Fig. 2 Demeanour of individuals** Stacked bar graphs illustrating the percentage of individuals showing different demeanours in high intermediate and low human flux zones.

**Fig. 3**

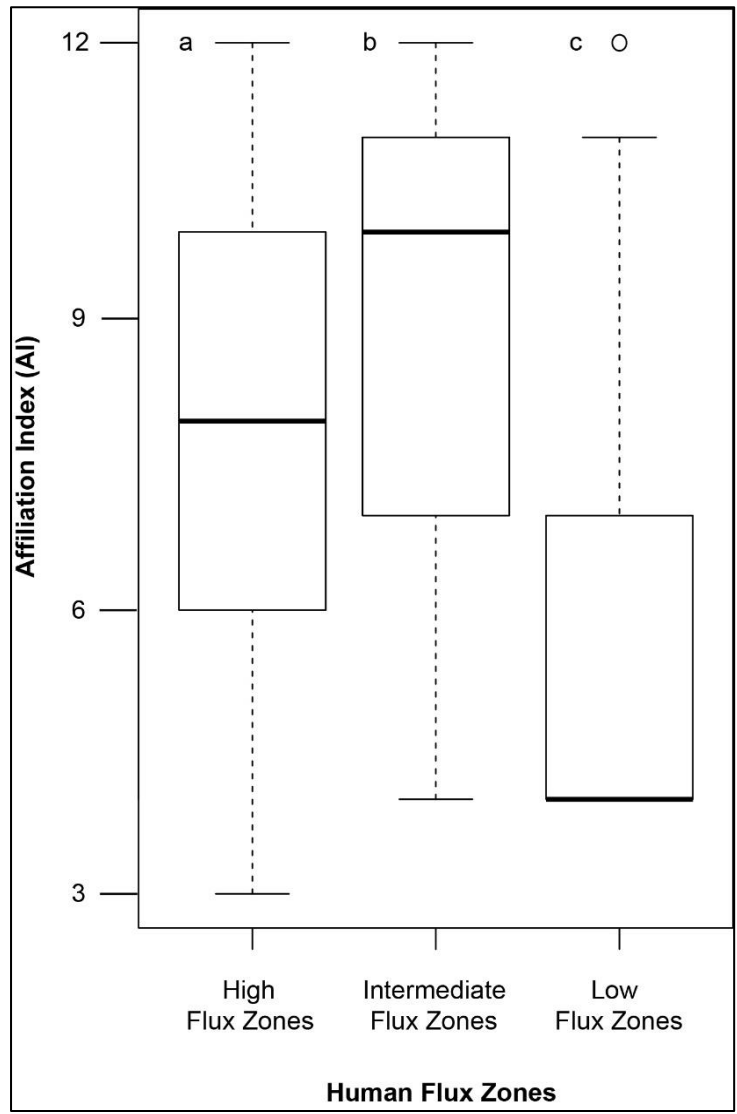

**Fig. 3 Affiliation Index** Box and whiskers plot illustrating the affiliation index values of individuals in high, intermediate and low human flux zones. Boxes represent interquartile range, horizontal bars within boxes indicate median values, and whiskers represent the upper range of the data. "a", "b" and "c" indicate significant differences.

# Supplementary Information

**Supplementary Table 1 - Effect of demeanour and human flux zones on approach.**

Generalised liner model showing an effect of LF zones on the approach response.

**Model:** Approach ~ Demeanour + Human flux zone, family = binomial (link = "logit")

|  | Estimate | Standard error | z- value | Pr(>|z|) |
|---|---|---|---|---|
| Coefficients | | | | |
| intercept | 19.909794 | 766.635007 | 0.026 | 0.979 |
| Demeanour_Aggressive | -20.603360 | 766.635974 | -0.027 | 0.979 |
| Demeanour_Anxious | -19.947245 | 766.634998 | -0.026 | 0.979 |
| Demeanour_Neutral | -18.050081 | 766.635024 | -0.024 | 0.981 |
| Zone_IF | 0.001257 | 0.286034 | 0.004 | 0.996 |
| Zone_LF | -1.534295 | 0.274754 | -5.584 | 2.35e-08 *** |

*** $P < 0.001$

**Supplementary Figure 1 – Bar graph showing mean density ± S.D of dogs in the HF, IF and LF zones.** 'a' and 'b' indicate significant differences.

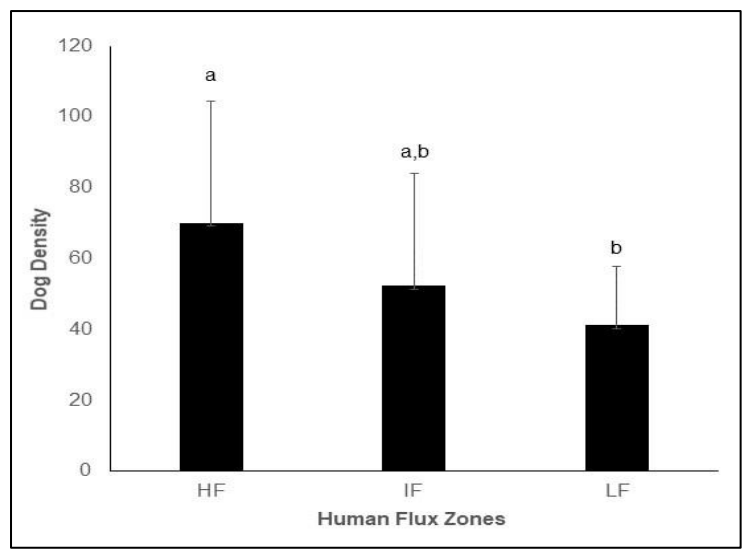

**Supplementary Figure 2 – Bar graph showing mean area ± S.D of the sampling sites in the three human flux zones.**

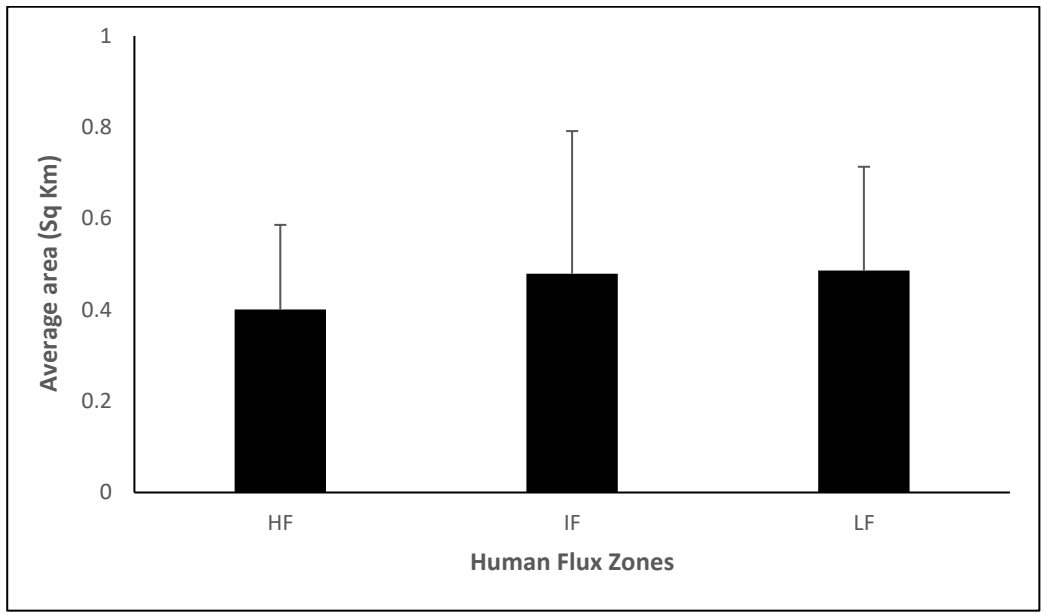

**Supplementary Figure 3 – Scatter plot showing correlation between RI values and dog abundance in the three human flux zones.** Solid black, green, and red dots indicate data of HF, IF, and LF zones respectively.

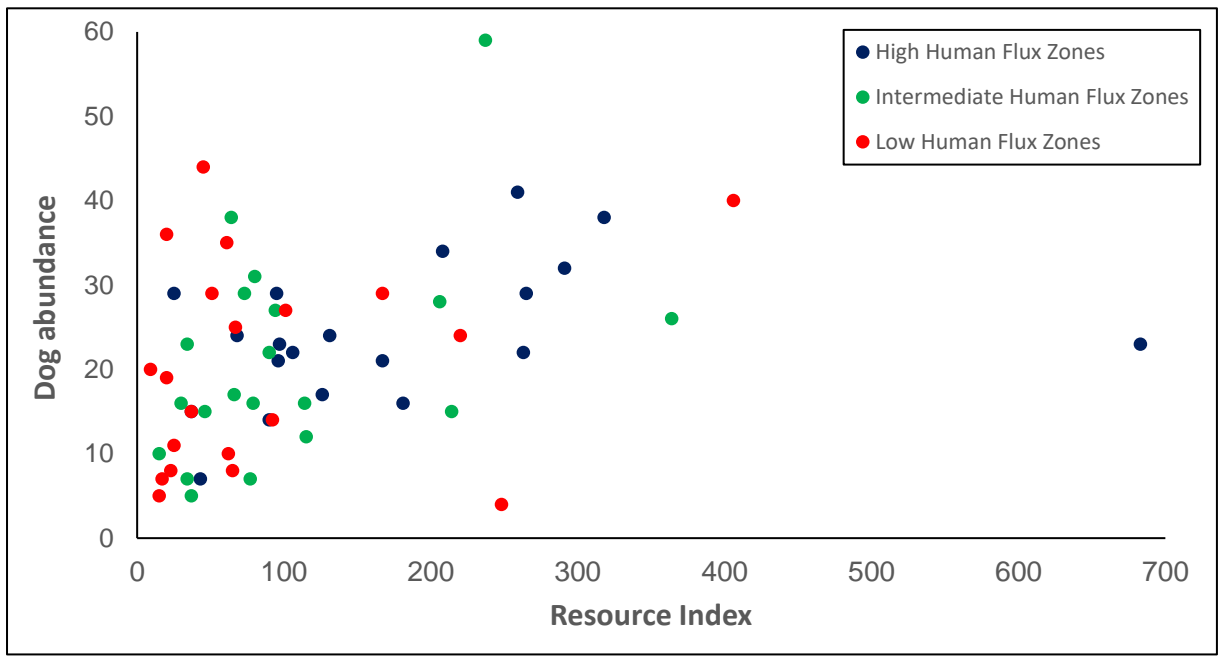

**Supplementary Table 2 – List of reasons describing liking and not liking dogs on streets.**

| Category | Reason |
|---|---|
| Liking dogs on streets | (i) Adorable, (ii) friendly and affectionate, (iii) loyal, (iv) living being like humans. |
| Not liking dogs on streets | (ii) Disturbance due to bark and howl, (iii) bite people and spread diseases, (ii) scatter garbage and create nuisance, (iii) chase humans and running vehicles. |

**Supplementary Table 3 - Effect of sex on approach.** Generalised liner model showing no effect of sex on the approach response.

**Model -** Approach ~ Sex, family = binomial (link = "logit")

|  | Estimate | Standard error | z- value | Pr(>|z|) |
|---|---|---|---|---|
| Coefficients | | | | |
| intercept | 0.5713 | 0.1198 | 4.769 | 1.86e-06 *** |
| Sex Male | -0.0926 | 0.1690 | -0.548 | 0.584 |

**Supplementary Table 4 - Effects of zones, Resource Index, and Perception Index on Affiliation Index.** Generalised liner mixed model showing effects of IF and LF zones on the Affiliation Index.

**Model –** Affiliation Index ~ Human flux zones + Resource Index + Perception Index + (1 | Sampling Site), family = Poisson ("log").

|  | Estimate | Standard error | z- value | Pr(>|z|) |
|---|---|---|---|---|
| Coefficients | | | | |
| intercept | 4.565e+00 | 1.444e-01 | 31.606 | < 2e-16 *** |
| Intermediate Human Flux Zone | 1.822e-01 | 4.706e-02 | 3.872 | 0.00010 *** |
| Low Human Flux Zone | -2.514e-01 | 4.420e-02 | -5.689 | 1.28e-08 *** |
| Resource Index (RI) | 9.035e-05 | 1.326e-04 | 0.681 | 0.495755 |
| Perception Index (PI) | -2.252e-03 | 1.276e-03 | -1.765 | 0.077583 . |

\*\*\* P < 0.001

**Supplementary Figure 4 –** Bar graph showing human movement (flux) ± S.D per minute in 24 sampling sites and categorisation of sites into HF, IF and LF zones.

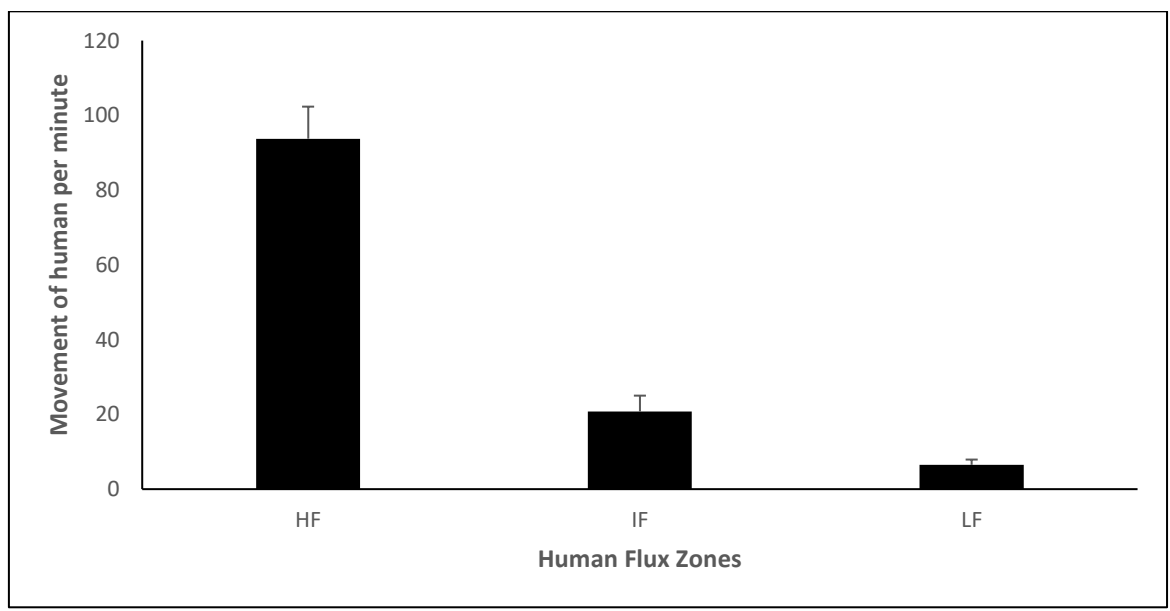

**Supplementary Figure 5 – Indian map showing the study areas (9 large cities). Experimentation was done in an around the 9 large cities.**

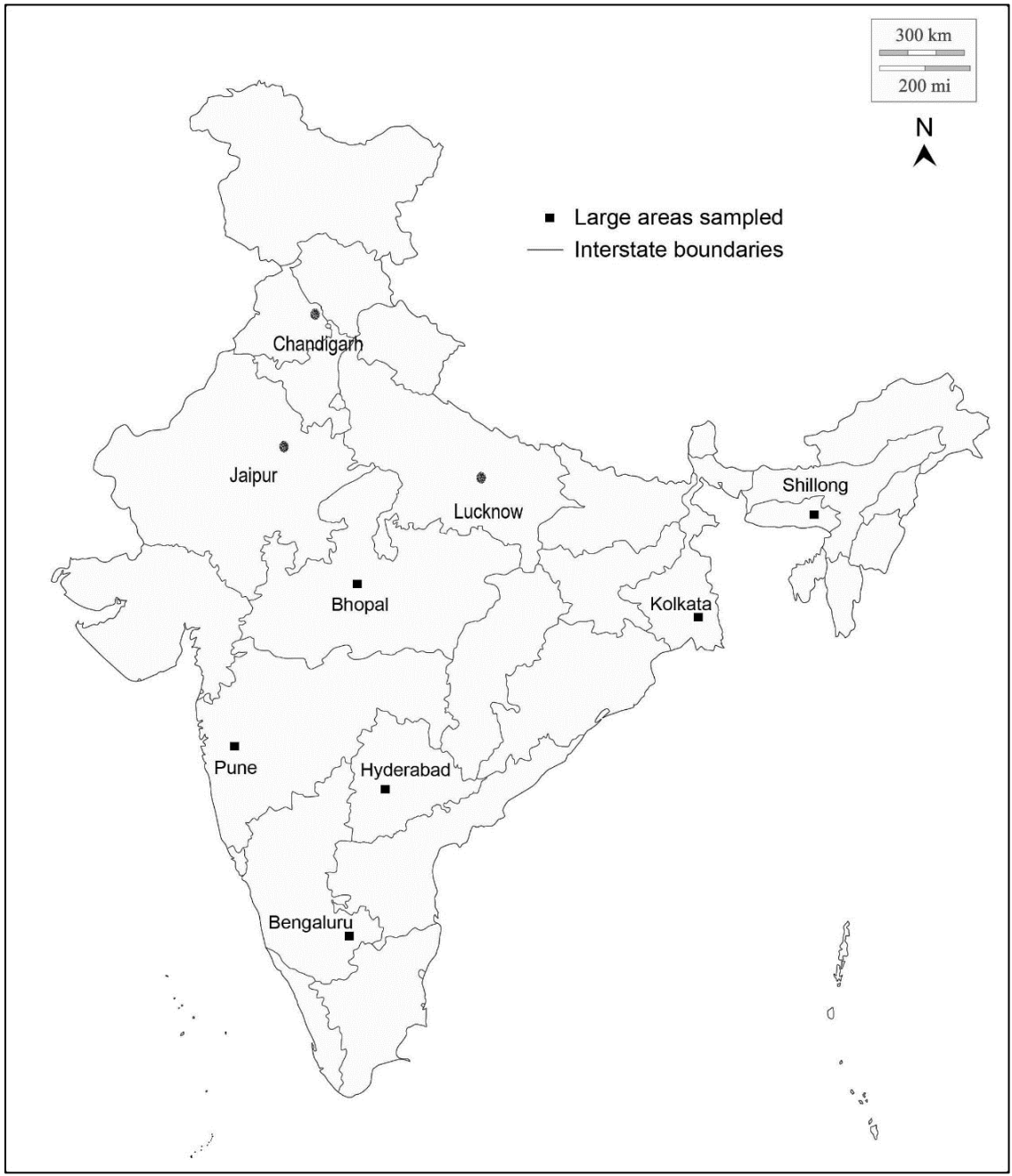

**Supplementary Table 5 – Distribution of all 60 sampling sites (in and around large city areas, Fig S5) and their categorisation into HF, IF and LF zones.**

| Large Area | Sampling Site | Observed human flux | Zone type |
|---|---|---|---|
| *Pune* | KEM Hospital Area | 84 | High |
| | Dhole Patil Road | 97 | High |
| | Shivajinagar | 90 | High |
| | Hanuman nagar | 81 | High |
| | Abhiruchi | 63 | High |
| | DSK Dhayari | 34 | Intermediate |
| | Benkarvasti | 29 | Intermediate |
| | Tridalnagar Co-op Society | 8 | Low |
| | Vimannagar | 3 | Low |
| | Nanded | 4 | Low |
| *Bengaluru* | Vivekanagar | 102 | High |
| | RT Nagar | 64 | High |
| | Marathalli | 82 | High |
| | St. Mary's Church area | 71 | High |
| | Bomanahalli | 32 | Intermediate |
| | Canara Bank Layout | 34 | Intermediate |
| | Sriramapura | 33 | Intermediate |
| | Akshayama Temple area | 7 | Low |
| | Silk Board area | 3 | Low |
| | Kudlu | 4 | Low |
| *Hyderabad* | Maruthinagar | 70 | High |
| | Matrix Hospital area | 61 | High |
| | Boduppal | 24 | Intermediate |
| | Ramireddy Nagar | 27 | Intermediate |
| | Chilkanagar | 4 | Low |
| | Dwarkamayee Colony | 9 | Low |
| | Kakatiya Park area | 6 | Low |
| *Kolkata* | Agarpara | 22 | Intermediate |
| | HB Town | 28 | Intermediate |
| | Kalyani 'A' Block | 30 | Intermediate |
| | Gayeshpur | 30 | Intermediate |
| | Jaguli | 29 | Intermediate |
| *Bhopal* | Habibganj | 68 | High |
| | Dev Mata Hospital area | 20 | Intermediate |
| | Pipalia Pende Kha | 26 | Intermediate |
| | Awadhpuri | 38 | Intermediate |
| | Srikrishnapuram | 7 | Low |
| | Bhauri | 7 | Low |
| | Amrawadh Khurdh | 8 | Low |
| *Shillong* | Nongmynsong | 94 | High |
| | Madarynting | 9 | Low |
| | NEHU area | 5 | Low |
| | Umpling | 8 | Low |
| *Lucknow* | Campbell Road | 89 | High |
| | Bara Imambara | 92 | High |
| | Nazirabad | 93 | High |
| | Alambagh | 77 | High |
| | Daulatganj | 30 | Intermediate |
| | Tikaitganj | 26 | Intermediate |
| | Daudnagar | 2 | Low |

| | Sekhpur | 6 | Low |
|---|---|---|---|
| *Chandigarh* | Maloya | 85 | High |
| | Balongi | 80 | High |
| | Sector 37 Market | 32 | Intermediate |
| | Daddu Majra | 27 | Intermediate |
| | Kansal | 7 | Low |
| | Kaimbwala | 5 | Low |
| *Jaipur* | Lal Bahadur Sastri Nagar | 84 | High |
| | Kundan nagar | 31 | Intermediate |
| | Taruchhaya nagar | 6 | Low |

**Supplementary Table 6 – Table summarizes the questions asked to people regarding their perception of free-ranging dogs.**

| Questions | Choice | | |
|---|---|---|---|
| Do you tolerate free-ranging dogs on streets? | Yes | No | Neutral |
| Do you feed free-ranging dogs? | Yes | No | - |
| Do you beat / harass / interact negatively with free-ranging dogs? | Yes | No | Neutral |

**Supplementary Table 7 – Behaviours coded and their definitions –**

| Behaviour | Definition |
|---|---|
| Approach | <u>PV phase</u> – The subject (dog) moved towards E and the distance between E and the subject was ≤ 0.3 m. |
| | <u>S phase</u> – The subject moved towards E and inspected (sniff or lick) or obtained the food. |
| No approach | The subject did not approach the task/experimenter in the PV or S phase. Distance between E and subject was > 0.3 m. |
| Latency | Time taken to approach (PV phase) or obtain / inspect the food after its provision on the ground (S phase). |

**Supplementary Table 8 – Components and their corresponding scores to build Human Affiliation Index**

| Human Affiliation Index (Component I + Component II + Component III) ||
|---|---|
| **Component I: Phase-based approach** | **Score** |
| Approach in PV Phase | 3 |
| Approach in S Phase | 2 |
| No approach | 1 |
| **Component II: Latency** | **Score** |
| Rapid (1 sec – 2 sec) | 5 |
| Fast (3 sec – 5 sec) | 4 |
| Moderate (6 sec – 10 sec) | 3 |
| Slow (11 sec onwards) | 2 |
| No latency (~ no approach) | 1 |
| **Component I: Demeanour** | **Score** |
| Affiliative | 4 |
| Neutral | 3 |
| Anxious | 2 |
| Aggressive | 1 |

**Supplementary Table 9 – Types of resources and assigned scores to construct Resource Index**

| Regular sources | Score |
|---|---|
| (i) Meat shop / Fish shop (Open meat / fish shops only) | 8 |
| (ii) Eatery / Restaurant (Roadside open eateries / restaurants) | 7 |
| (iii) Direct feeding by human (provision of food directly by humans e.g. leftover food from human households) | 6 |
| (iv) Open Garbage Dump (minimum 2 Sq m in size, consisting of wet and dry garbage including leftover foods by human. It can accommodate more than 2 individuals) | 5 |
| (v) Tea stalls (Open tea stalls) | 4 |
| **Irregular sources** | **Score** |
| (vi) Direct feeding by human (occasional provisioning of food) | 3 |
| (vii) Grocery / Sweet shops / Bakery | 2 |
| (viii) Household or eatery bins without covers (less than or equal to 0.5 Sq m in size consisting of wet and dry garbage, can accommodate up to 2 individuals) | 1 |

**Supplementary Table 10 – Questions, response and corresponding scores to build Perception Index**

| Question | Response | Score |
| --- | --- | --- |
| Tolerance of free-ranging dogs on streets | Yes | 3 |
| | Neutral | 2 |
| | No | 1 |
| Feeding free-ranging dogs | Yes | 2 |
| | No | 1 |
| Initiate negative behaviour towards free-ranging dogs | No | 3 |
| | Neutral | 2 |
| | Yes | 1 |

**Figure 6 – Images showing samples of the three types of zones: (a) High Human Flux Zone (b) Intermediate Human Flux Zone and (c) Low Human Flux Zone.**

Photo courtesy: (a) Shubhra Sau, (b) Debottam Bhattacharjee, and (c) Rohan Sarkar.

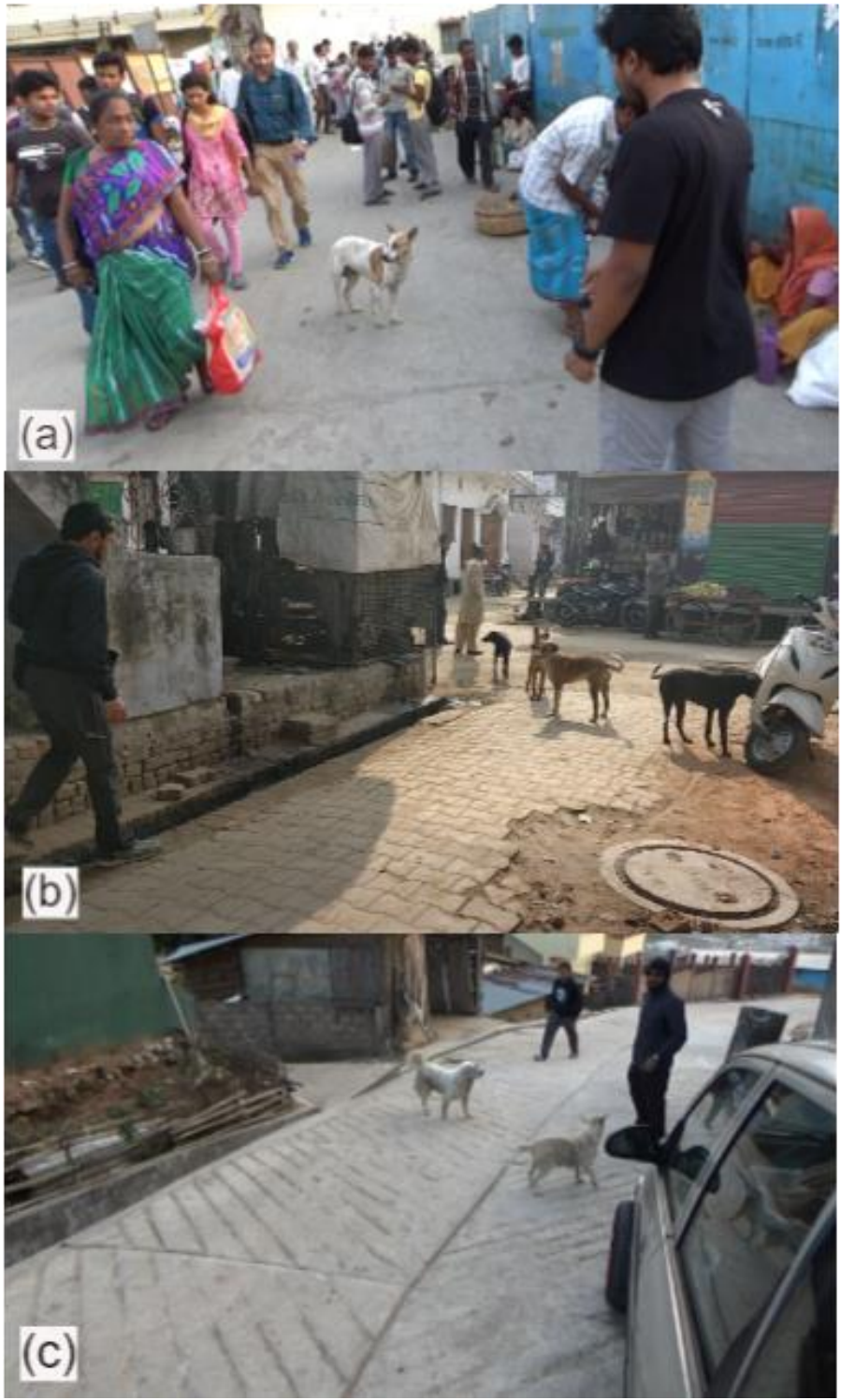

**Supplementary Movie 1 –** Video clip demonstrating a successful Positive vocalisation (PV) phase.

**Supplementary Movie 2 –** Video clip demonstrating a failed PV phase.

**Supplementary Movie 3 –** Video clip demonstrating a successful Stimulus (S) phase.